\documentclass[12pt]{iopart}
\usepackage{graphicx}% Include figure files
\usepackage{dcolumn}% Align table columns on decimal point
\usepackage{bm}% bold math

\begin{document}

\title[Two-Channel Exclusion Processes]{Two-Channel Totally Asymmetric Simple Exclusion Processes}
\author{Ekaterina Pronina and Anatoly B. Kolomeisky}
\address{Department of Chemistry, Rice University, Houston, TX 77005-1892}

\begin{abstract}
Totally asymmetric simple exclusion processes, consisting of two coupled parallel lattice chains with particles  interacting with hard-core exclusion and moving along the channels and between them, are considered. In the limit of strong coupling between the channels, the particle currents, density profiles and a phase diagram  are calculated exactly by mapping the system into  an effective one-channel totally asymmetric exclusion model. For intermediate couplings, a simple approximate theory, that describes the particle dynamics in  vertical clusters of two corresponding parallel sites  exactly and neglects the correlations between different vertical clusters, is developed. It is found that, similarly to the case of one-channel totally asymmetric simple exclusion processes, there are three stationary state phases, although the phase boundaries and stationary properties strongly depend on inter-channel coupling. An extensive computer Monte Carlo simulations fully support the theoretical predictions.   
\end{abstract}


\ead{tolya@rice.edu}

\submitto{\JPA}

\maketitle

\section{Introduction}

In recent years asymmetric simple exclusion processes (ASEPs) have become an important tool of investigation for many  processes in chemistry, physics and biology \cite{derrida98,schutz}.  ASEPs have been applied successfully  to understand the kinetics of biopolymerization \cite{macdonald68},  polymer dynamics in dense medium \cite{schutz99}, diffusion through membrane channels \cite{chou98}, gel electrophoresis \cite{widom91}, dynamics of motor proteins moving along rigid filaments \cite{klumpp03},  and the kinetics of synthesis of proteins \cite{lakatos03,shaw03,kolomeisky04}.

ASEPs are one-dimensional  lattice models where particles interact only with  hard-core  exclusion potential. Each lattice site can  be occupied by a particle or it can be empty. In the simplest totally asymmetric simple exclusion process (TASEP), where the particles can only move in one direction, the dynamic rules are the following. The particle at site $i$ can jump forward to the site $i+1$ with the rate 1 if the target site is empty. The particle can enter the lattice with the rate $\alpha$, provided the first site is unoccupied, and it can also leave the system with the rate $\beta$. Although these dynamic rules are very simple, they lead to a very rich and complex dynamic phase behavior. There are non-equilibrium phase transitions  between the stationary states of the system, induced by boundary processes, that have no analogs in equilibrium systems \cite{derrida98,schutz,kolomeisky98}.

The coupling of ASEPs with different equilibrium and non-equilibrium processes have led to many  unusual and unexpected phenomena. An introduction of a single irreversible detachment process to the bulk of the system have resulted in significant changes in phase diagram \cite{mirin03}. The coupling of ASEPs with equilibrium Langmuir kinetics at each site produced  unusual phenomena of localized density shocks \cite{parmegianni03,popkov03,evans03,levine04}. However, the complexity of phase transitions in  ASEPs can be reasonable well explained and understood by applying a phenomenological domain-wall theory \cite{schutz,kolomeisky98}.

Most investigations of ASEPs concentrate on one-channel systems where particles can only move along one lattice chain. However, the description of  many real phenomena would be more adequate if  parallel-chain asymmetric exclusion processes are used. For example, motor proteins kinesins can move along the parallel protofilaments of microtubules, and there are no restrictions for them to jump between these protofilaments \cite{howard}. Parallel-chains ASEPs have been considered earlier \cite{popkov01,ps03,popkov04a}. However, in these investigations the coupling between different chains was indirect, i.e., hopping between the chains was forbidden. The aim of the present paper is to investigate two-chain asymmetric exclusion processes, where particles can move between the lattice channels.  We investigate this system by using a simple approximate model for intermediate couplings and by mapping it to the exactly solved one-channel TASEP in the limits of strong coupling and no coupling. In addition, extensive computer Monte Carlo simulations are performed in order to validate theoretical predictions.

The paper is organized as follows. In section 2 we give a detailed description of the model, discuss known results for one-channel ASEPs, solve exactly the stationary state properties of the system in the limit of strong coupling, and develop an approximate theory for two-channel asymmetric exclusion processes for intermediate couplings. Then in section 3 we present and discuss Monte Carlo simulations results. Finally, we summarize and conclude in section 4.

\section{Theoretical Description of Two-Channel TASEP}

\subsection{Model}

\begin{figure}[tbp]
\centering
\includegraphics[clip=true]{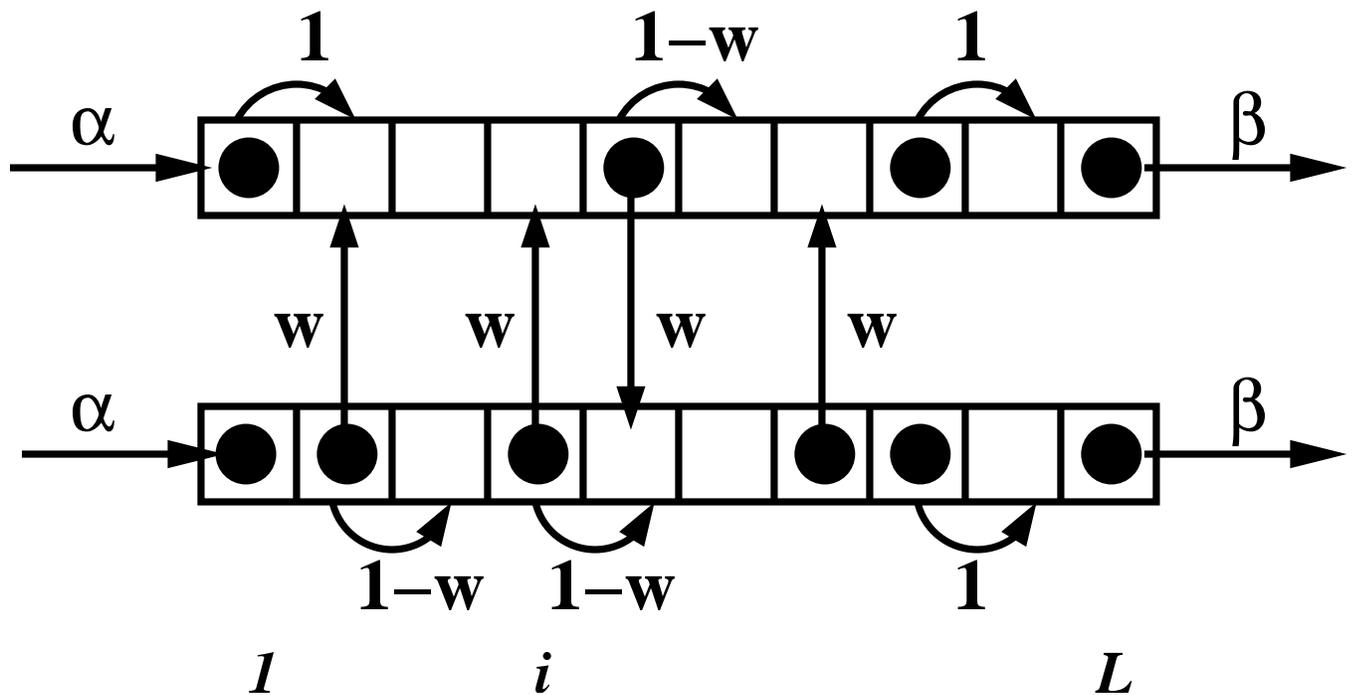}
\caption{Schematic view of the model for two-channel TASEP. Allowed transitions are shown by arrows. Inter-channel hopping rates are equal to $w$.  The transition rates to move along the channel are $1-w$, if there is no particle at the same site in another channel, otherwise it is equal to 1. Entrance and exit rates are $\alpha$ and $\beta$ (or $\beta(1-w)$ if the last cluster is half-empty), respectively, and are the same in both channels.}
\label{fig1}
\end{figure}

We consider identical particles moving on two parallel one-dimensional lattices, each lattice has $L$ sites, as shown in Fig. \ref{fig1}. Every site on both lattices can be either empty or it can be occupied by  no more than one particle.  In the bulk of the system the dynamic rules are the following. A particle at site $i$ can hop up or down to the same site $i$ on the other lattice chain with the rate $0 \le w \le 1$, if that site is empty. The particle can also move from left to right along the same lattice to site $i+1$, if this site is available. However, this transition rate depends on occupation of site $i$ at neighboring channel. If there is no particle at that site, the rate is equal to $1-w$, otherwise the particle jumps with the rate 1 (see Fig. \ref{fig1}). It means that the full transition rate of leaving site $i$, to go forward or up/down, is always equal to 1. Particles can also enter the system with the rate $\alpha$ if any of the first sites in either  channel are empty. When a particle reaches site $L$, it can exit with the rate $\beta$, when both last sites are occupied, or with the rate $\beta(1-w)$ if there is no vertical neighbor at the other lattice chain.  
 
When a vertical transition rate $w=0$,  we have two uncoupled single-channel totally asymmetric simple exclusion processes for which an exact description of phase diagram, density profiles and particle currents is known \cite{derrida98,schutz}. In this case there are three steady-state phases, determined by the processes at the boundaries or in the bulk of the system.  For $\alpha >1/2$ and $\beta > 1/2$, the particle dynamics is specified by the processes in the bulk of the system, and we have a maximal-current phase with the following stationary current and bulk density,
\begin{equation}
J_{MC}=1/4, \quad \rho_{bulk,MC}=1/2.
\end{equation}
The entrance into the system determines the overall particle dynamics for $\alpha < \beta$ and $\alpha < 1/2$. In this case the system can be found in a low-density phase with the current and bulk density given by
\begin{equation}
J_{LD}=\alpha (1-\alpha), \quad \rho_{bulk,LD}= \alpha.
\end{equation}
When the exit from the system limits the overall dynamics, which takes place for $\beta < \alpha$ and $\beta < 1/2$, the system exists in a high-density phase. In this case the current and bulk densities are equal to
\begin{equation}
J_{HD}=\beta (1-\beta), \quad \rho_{bulk,HD}= 1-\beta.
\end{equation}

\begin{figure}[tbp]
\centering
\includegraphics[clip=true]{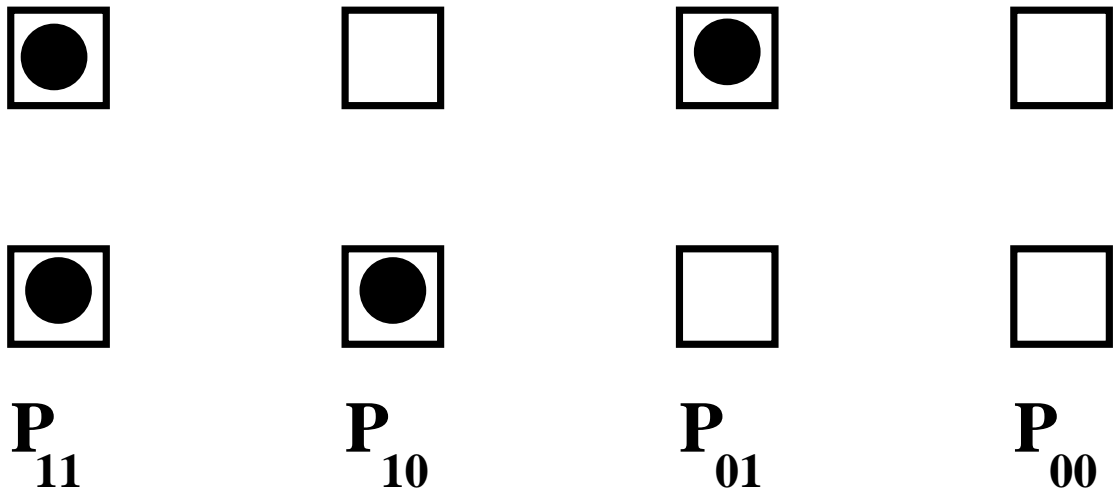}
\caption{Four possible states for a vertical cluster of lattice sites. $P_{11}$, $P_{10}$, $P_{01}$ and $P_{00}$ are the corresponding probabilities for each state.}
\label{fig2}
\end{figure}

For vertical transition rates $w>0$, the particle dynamics in both channels depend on each other, and any successful theoretical description should be able to account for this coupling. Let us consider a  cluster of two vertical sites $i$. There are four possible states for this cluster, as shown in Fig. \ref{fig2}.$P_{11}^{(i)}$ is defined as a probability to find a cluster with both lattice sites occupied. Then  $P_{10}^{(i)}$ and $P_{01}^{(i)}$ are the probabilities that only lower or upper channel site is occupied (see Fig. \ref{fig2}). Finally,  $P_{00}^{(i)}$ is a probability that both sites in the  cluster are empty. The normalization condition  for these probabilities gives us
\begin{equation} \label{norm}
P_{11}^{(i)}+P_{10}^{(i)}+P_{01}^{(i)}+P_{00}^{(i)}=1.
\end{equation}

In the bulk of the system at stationary state, it is reasonable to expect that these probabilities are independent of the position of the vertical cluster due to translational symmetry. In what follows, we  omit the superscript $i$ for bulk values of these probabilities since only steady-state processes will be considered. Also, because  both channels are equivalent, we expect that $P_{10}^{(i)}=P_{01}^{(i)}$.  Then the bulk density at each lattice chain, which is also independent of the position on the lattice in stationary state limit, is given by
\begin{equation}\label{density}
\rho=P_{11}+P_{10}.
\end{equation}
Thus the overall dynamics of the system can be fully described in terms of these probabilities.

\subsection {Strong-Coupling Limit}

In the limit of strong coupling ($w=1$), the dynamics of the system simplifies significantly, because at large times any vertical cluster cannot exist in  a configuration, where both sites are empty, i.e., $P_{00}=0$. This can be seen from the following arguments. A vertical cluster in  state $\{00\}$ could only be obtained  by moving particle along the corresponding lattice chain if the previous state of the same cluster was $\{10\}$ or $\{01\}$. However, the rate for this transition in the limit of strong coupling is equal to $1-w=0$, and the configuration $\{00\}$ can never be reached for any vertical cluster. Thus, in this system there are only two types of clusters: fully filled and half-filled. 

One can think of filled clusters as  new effective ``particles'' and half-filled clusters then can be viewed as new effective ``holes.'' The two-channel TASEP in this limit can be mapped into the one-channel totally asymmetric exclusion process with the effective entrance rate $\alpha$ and the effective exit rate $2 \beta$. In the bulk of the system an effective new particle jumps to the right with rate 1, if this move is allowed. The factor 2 in  the effective exit rate is due to the fact that there are two ways of leaving the system from the filled cluster at last site, namely, by moving particles from the upper or lower lattice channels. 

The exact density profiles, particle currents and phase diagram for this effective one-channel TASEP system are known \cite{derrida98,schutz}, and it allows us to calculate exactly the properties of the original two-channel TASEP in the limit of strong coupling.  The steady-state particle current of the effective one-channel model $J^{*}$ is related to the particle current per channel $J$ of the original two-channel model in the following way,
\begin{equation}
J=J^{*}/2,
\end{equation}
while using Eq. \ref{density} we obtain the corresponding relation for bulk density profiles in two-channel TASEP,
\begin{equation}
\rho=\rho^{*}+(1-\rho^{*})/2=(1+\rho^{*})/2,
\end{equation}
where $\rho^{*}$ is the effective density for one-channel asymmetric exclusion model.

Thus we conclude that there are 3 stationary-state phase for two-channel TASEP in the limit of strong coupling. When $\alpha < 2 \beta$ and $\alpha <1/2$, the system can be found in the low-density phase with the following properties,
\begin{equation}
P_{11,LD}=\alpha, \quad P_{10,LD}=(1-\alpha)/2, \quad J_{LD}=\alpha(1-\alpha)/2, \quad \rho_{bulk,LD}=(1+\alpha)/2.
\end{equation}
The conditions $\alpha >2 \beta$ and $\beta < 1/4$ specify the  high-density phase. In this case the steady-state properties are given by 
\begin{equation}
P_{11,HD}=1-2 \beta, \quad P_{10,HD}=\beta, \quad J_{HD}=\beta(1-2 \beta), \quad \rho_{bulk,HD}=1-\beta.
\end{equation}
In the case when bulk dynamics is rate-limiting, we have the maximal-current phase with the following parameters,
\begin{equation}
P_{11,MC}=1/2, \quad P_{10,MC}=1/4, \quad J_{MC}=1/8, \quad \rho_{bulk,MC}=3/4.
\end{equation}

\subsection{Approximate Solutions for Intermediate Couplings}

For intermediate couplings, $0<w<1$, any vertical cluster can exist in all four possible states and we cannot map the two-channel TASEP into the effective one-channel model. Some reasonable approximations are needed in order to calculate the steady-state properties of the system. 

Assume that a state of a given vertical cluster is independent of states of its neighbors. It means that the cluster dynamics is considered in a mean-field description. Then the probability of finding the cluster at the position $i$ with both sites occupied changes in time as given by
\begin{equation}
\frac{d P_{11}}{dt} = P_{11}(P_{10}+P_{01})+2(1-w)P_{10}P_{01}-2P_{11}P_{00}-P_{11}(P_{10}+P_{01}),
\end{equation}
where the first term corresponds to the probability density flux from  sites $i-1$, the second term describes the changes at both sites $i$ that lead to the formation of the cluster $\{11\}$, and two negative terms represent the density flux leaving from sites $i$. Applying the symmetry relation $P_{10}=P_{01}$ at large times simplifies this expression  into 
\begin{equation} \label{master}
0 = (1-w)P_{10}^{2}-P_{11}P_{00}.
\end{equation}
This equation, with the help of the normalization condition (\ref{norm}), yields 
\begin{equation} \label{P10expression}
P_{10}=\frac{-P_{11}+\sqrt{P_{11}^{2}+(1-w) P_{11}(1-P_{11})}}{1-w}.
\end{equation}  
Such rather complex relation can take a simpler form for two limiting cases - for two-channel TASEP without coupling  and in the strong-coupling limit. For $w=0$ we obtain 
\begin{equation}
P_{10}=\sqrt{P_{11}}-P_{11},
\end{equation} 
while for $w=1$ it transforms into
\begin{equation}
P_{10}=\frac{1-P_{11}}{2},
\end{equation} 
in agreement with exact calculations presented above [see Eqs. (8-10)].

The  particle current per channel can be written as
\begin{equation} \label{current}
J=\left[P_{11}+(1-w)P_{10}\right](1-P_{10}-P_{11}). 
\end{equation} 
In this expression the first multiplier gives the probability to find a vertical cluster in  configurations, from which the particle can hop to the right, while the second multiplier gives the probability that the site ahead is available. Using the relation (\ref{P10expression}) we can express the particle current only in terms of one variable - the density of vertical clusters $P_{11}$,
\begin{equation} \label{current1}
J=\sqrt{P_{11}(1-w+w P_{11})}\frac{1-w+wP_{11}-\sqrt{P_{11}(1-w+wP_{11})}}{1-w}. 
\end{equation}  
Again,  simpler relations can be obtained  for two limiting cases of zero coupling and strong coupling. The particle current in the case of $w=0$ is given by
\begin{equation}   
J=\sqrt{P_{11}}\left(1-\sqrt{P_{11}}\right),
\end{equation}
while for $w=1$ we conclude that
\begin{equation}   
J=P_{11}(1-P_{11})/2.
\end{equation}
These relations also agree with exact results discussed above.

We expect that, similarly to the case of zero inter-channel coupling or in the strong-coupling limit, there are three possible stationary phases. The conditions for a maximal-current phase can be specified by computing  a maximum of the particle current  as a function of density of $\{11\}$ clusters, i.e., $\frac{\partial J}{\partial P_{11}}=0$. It leads to the following equation,
\begin{equation}   
\frac{\left[ 1.5 w \sqrt{P_{11}(1-w+wP_{11})} +\frac{(1-w+wP_{11})^{1.5}}{2\sqrt{P_{11}}}-1+w-2wP_{11} \right]}{1-w}=0.
\end{equation}
For any value of inter-channel coupling $w$ this equation can always be solved  exactly numerically, and we can obtain the value of the density of filled vertical clusters in the maximal-current phase. Because this equation can have more than one real solution, we should always choose the physically reasonable solution which gives $1/4 < P_{11,MC} <1/2$. Utilizing Eqs. (\ref{density}), (\ref{P10expression}) and (\ref{current1}), the stationary properties of this phase can be easily determined. For example, for $w=0.5$ the calculations show that
\begin{equation}  
P_{11,max}\simeq 0.317, \quad P_{10,max} \simeq 0.280, \quad \rho_{bulk,max} \simeq 0.597, \quad J_{max} \simeq 0.184.
\end{equation}

In the low-density phase the entrance rate $\alpha$ limits the  overall particle dynamics, and the expression for the current per channel is given by 
\begin{equation}
J_{LD}=\alpha  (1-P_{11}-P_{10}),
\end{equation} 
or, using Eq. (\ref{P10expression}), it can be written as
\begin{equation}\label{current_LD}
J_{LD}=\alpha  \frac{1-w+wP_{11}-\sqrt{P_{11}(1-w+wP_{11})}}{1-w}.
\end{equation} 
Comparing this equation  with the expression for the particle current in the bulk of the system [see Eq.(\ref{current1})], it can be shown that 
\begin{equation}
\alpha=\sqrt{P_{11}(1-w+w P_{11})},
\end{equation}
from which the value of density of $\{11\}$ clusters in the low-density phase can be obtained as follows,
\begin{equation}\label{P11expression}
P_{11,LD}=\frac{-1+w+\sqrt{(1-w)^{2}+4w\alpha^{2}}}{2w}.
\end{equation}
Applying again Eq. (\ref{P10expression}), we derive the expression for the density of $\{10\}$ clusters,
\begin{equation}
P_{10,LD}=\frac{\alpha}{1-w} + \frac{1-\sqrt{1+w \left(\frac{2\alpha}{1-w}\right)^{2}}}{2w}.
\end{equation}
These equations allow us to calculate the bulk density as $\rho=P_{11}+P_{10}$, and also, substituting  Eq. (\ref{P11expression}) into Eq. (\ref{current_LD}), the particle current per channel can be expressed as
\begin{equation}\label{current_LD1}
J_{LD}=\frac{\alpha}{2} \left[ 1-\frac{2\alpha}{1-w}+\sqrt{1+4w \left(\frac{\alpha}{1-w}\right)^{2}} \right].
\end{equation}  

When  exit dynamics determines the overall behavior of the system we have a high-density phase. The particle current per channel is given by
\begin{equation}
J_{HD}=\beta  \left[P_{11}+(1-w)P_{10}\right],
\end{equation} 
and, after applying Eq. (\ref{P10expression}), it transforms into
\begin{equation}\label{current_HD}
J_{HD}=\beta \sqrt{P_{11}(1-w+w P_{11})}.
\end{equation} 
Again comparing this expression with the particle current in the bulk, as given by  Eq.(\ref{current1}), we derive the density of filled vertical clusters,
\begin{equation}\label{P11}
P_{11,HD}=1-\beta-\frac{1-\sqrt{1-4w\beta(1-\beta)}}{2w}.
\end{equation}
Substitution of  it into Eq. (\ref{P10expression}) allows us to compute the density of $\{10\}$ clusters,
\begin{equation}
P_{10,HD}=\frac{1-\sqrt{1-4w\beta(1-\beta)}}{2w}.
\end{equation}
Then the bulk density in each channel is equal to $\rho_{HD}=P_{11,HD}+P_{10,HD}=1-\beta$. It is interesting to note that the bulk density in this phase is always independent of inter-channel coupling and depends only on exit rate, although the densities of filled and half-filled clusters strongly depend on $w$. The expression for the particle current can be derived from Eqs. (\ref{current_HD}) and (\ref{P11}),
\begin{equation}\label{current_HD1}
J_{HD}=\frac{\beta}{2}\left[1-2\beta+\sqrt{1-4w\beta(1-\beta)}\right].
\end{equation} 

\begin{figure}[tbp]
\centering
\includegraphics[clip=true]{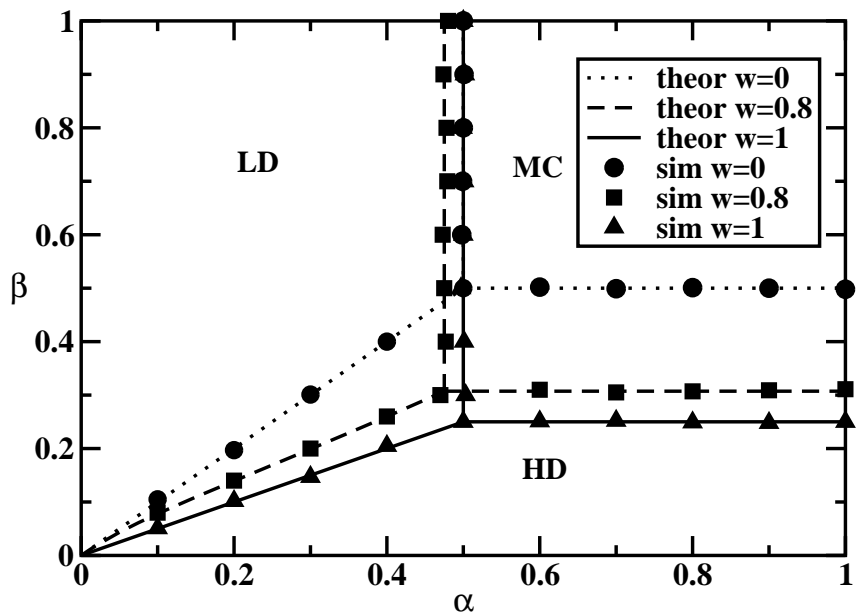}
\caption{Phase diagrams for two-channel TASEPs for different inter-channel couplings. Symbols correspond to Monte Carlo computer simulations, lines are our theoretical predictions.}
\label{fig3}
\end{figure}

There are two types of phase transitions in the system. The boundary between the low-density and high-density phase specifies the first-order phase transition with jumps in bulk densities, while the transition between the maximal-current and low-density or high-density phases  is continuous. The low-density and high-density phases coexist when the particle currents in each phase are equal, i.e., $J_{HD}=J_{LD}$. From Eqs. (\ref{current_LD1}) and (\ref{current_HD1}) we obtain,
\begin{equation}
\frac{\alpha}{2} \left[1-\frac{2\alpha}{1-w}+\sqrt{1+4w\left(\frac{\alpha}{1-w}\right)^{2}} \right]=\frac{\beta}{2}\left[1-2\beta+\sqrt{1-4w\beta(1-\beta)}\right],
\end{equation} 
which determines the phase boundary curve in $\{\alpha, \beta\}$ coordinates. One can observe that, in contrast to the systems with zero coupling or in the limit of strong coupling, the phase boundary here is not a linear function of parameters. Similar arguments can also be used to find the boundaries between the low-density or high-density and the maximal-current phases. Theoretically calculated phase diagrams for different inter-channel couplings are presented in Fig. \ref{fig3}.

\section{Monte-Carlo Simulations and Discussions}

Our approximate theory treats exactly the particle dynamics inside vertical clusters, however, interactions between the clusters are accounted for in an approximate mean-field fashion. Theoretical predictions from this approximate method agree with exact results in the limiting cases of zero inter-channel coupling and strong coupling. To check the overall validity of our approach for intermediate couplings we performed extensive Monte Carlo computer simulations.

Our theoretical arguments are valid only in the thermodynamic limit, i.e., $L \rightarrow \infty$. However, in our simulations the lattice of  size  $L=100$ was used, and we checked that for larger lattices there is no difference with reported results. The density profiles and particle currents in our simulations were calculated by averaging over $10^{8}-10^{10}$ Monte Carlo steps, although first $5 \% $ of total number of steps were neglected to ensure that the system has reached the stationary state. Boundaries in phase diagrams were determined by comparing density profiles and changes in the particle current. 

\begin{figure}[tbp]
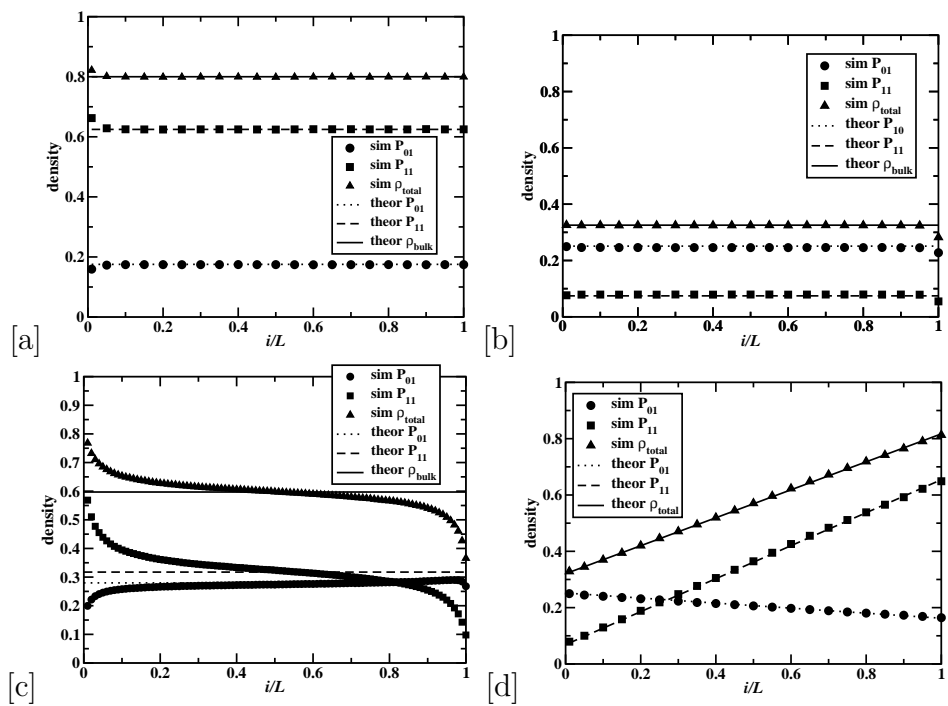

\centering
[a]\includegraphics[clip=true]{Fig4a.eps}
[b]\includegraphics[clip=true]{Fig4b.eps}
[c]\includegraphics[clip=true]{Fig4c.eps}
[d]\includegraphics[clip=true]{Fig4d.eps}
\caption{Density profiles for inter-channel coupling $w=0.5$:  a) in the high-density phase with $\alpha=0.8$ and $\beta=0.2$; b) in the low-density phase with $\alpha=0.2$ and $\beta=0.8$; c) in the maximal-current phase with $\alpha=0.8$ and $\beta=0.8$; and d) at the boundary between the low-density and the high-density phases for $\alpha=0.2$ and $\beta=0.2$.  Symbols correspond to Monte Carlo computer simulations results,  lines describe theoretically calculated  bulk values for $P_{10}$, $P_{11}$ and density per channel, respectively. Error bars, determined from standard deviations for simulations, are smaller than the size of the symbols.}\label{fig4}
\end{figure}

Phase diagrams and density profiles calculated from Monte Carlo simulations are presented in Figs. \ref{fig3} and \ref{fig4}. The agreement with theoretically calculated values is excellent. It should be noted that the phase boundary between the low-density and the high-density phases for intermediate coupling (for $0<w<1$) deviates slightly from linear relation, in agreement with theoretical predictions. That can be seen by analyzing the slope of the curve. It equals to 1 at small $\alpha$ and $\beta$ and slowly decreases for higher values of parameters, as shown in Fig. \ref{fig3}.  The overall effect of inter-channel couplings is the decrease in a phase volume for the high-density phases. Note, that although we presented only theoretical bulk values for densities of clusters and particle densities (see Fig. \ref{fig4}), the approximate theory allows us also to calculate the full density profiles.

\begin{figure}[tbp]
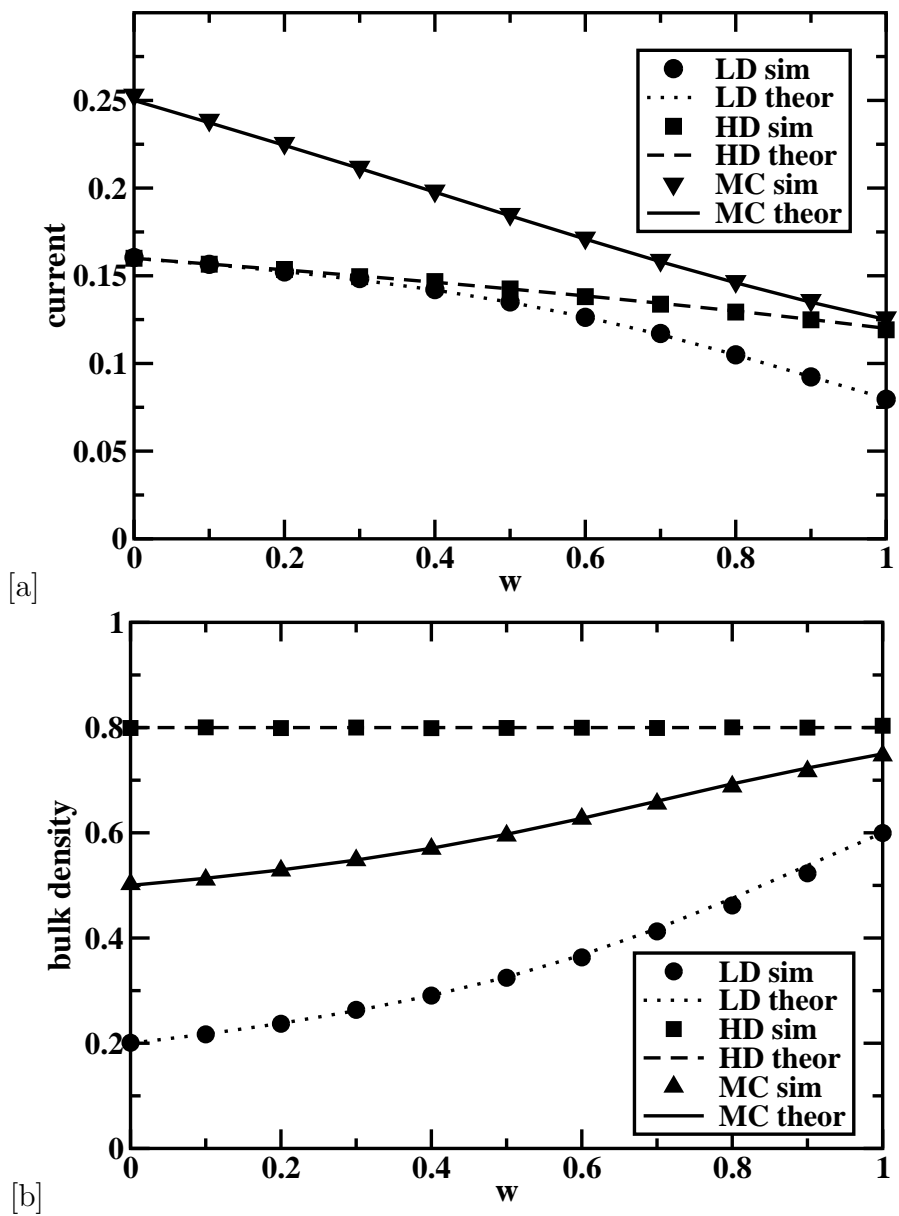

\centering
[a]\includegraphics[clip=true]{Fig5a.eps}
[b]\includegraphics[clip=true]{Fig5b.eps}
\caption{Particle currents per channel and bulk densities as a function of inter-channel coupling. The low-density phase is specified by $\alpha=0.2$ and $\beta=0.8$; the high-density phase is given by $\alpha=0.8$ and $\beta=0.2$; and the maximal-current phase is for $\alpha=0.8$ and $\beta=0.8$. Symbols correspond to Monte Carlo computer simulations, lines are our theoretical predictions.}
\label{fig5}
\end{figure}

\begin{figure}[tbp]
\centering
\includegraphics[clip=true]{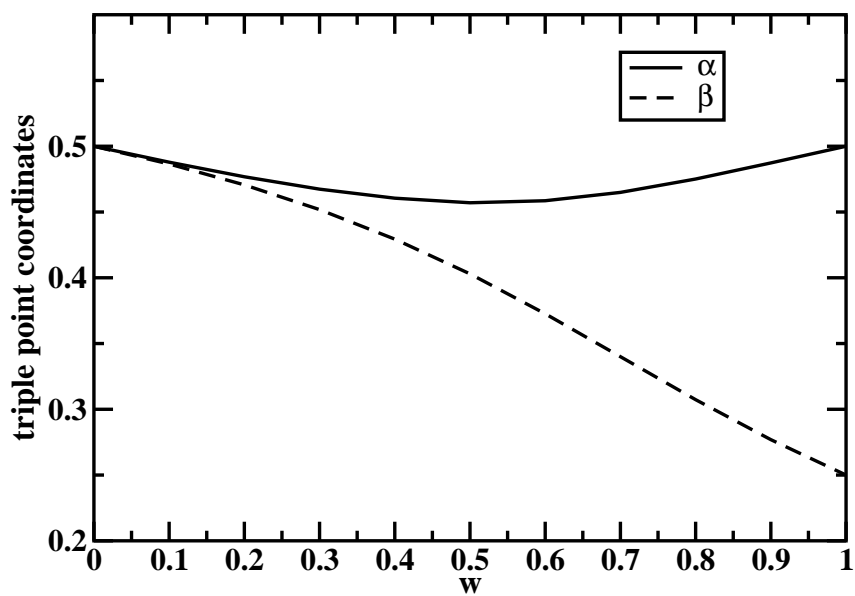}
\caption{The coordinates of triple points for different inter-channel couplings. Monte Carlo computer simulations results coincide with the theoretical calculations, and thus are not shown.}
\label{fig6}
\end{figure}

It is interesting also to analyze  the effect of inter-channel particle transitions on the dynamics of two-channel TASEPs. As shown in Fig. \ref{fig5}, the increase in inter-channel coupling generally decreases the particle current per channel, while the values of bulk densities do not change (for the high-density phase) or go up. The strongest effect is for the maximal-current and the low-density phases, while the influence of coupling on the high-density phase is quite minimal. That can be easily understood by recalling what processes determine the different steady-state phases. The entrance processes and bulk processes are strongly affected by inter-channel particle transitions, while the exit processes do not depend much on this coupling.  The inter-channel coupling has also a peculiar effect on a triple point in phase diagram, where all three phases coexist - see Fig. \ref{fig6}. The increasing inter-channel current decreases $\beta$-coordinate of the triple point, while the effect on $\alpha$-coordinate is non-monotonous.

\section{Summary and Conclusions}

We investigated two-channel totally asymmetric simple exclusion processes with the possibility of particle transitions  between the lattice chains. When there is no coupling, $w=0$, the system can be viewed as two independent one-channel TASEPs, for which exact solutions are known. In the limit of strong coupling, $w=1$, the exact description of particle dynamics is obtained by mapping the two-channel system into an effective one-channel TASEP with known stationary properties. 

The two-channel TASEPs with intermediate couplings are analyzed with the help of an approximate theoretical approach, in which all stationary properties are  obtained analytically. In our approach the particle dynamics inside vertical clusters of corresponding lattice sites is considered exactly, while the correlations between different clusters are neglected. The results of this approximate method are in excellent agreement with computer Monte Carlo simulations. Our theoretical and computational results indicate that  the inter-channel currents  have a  strong effect on steady-state properties of the system. Increasing the coupling lowers the particle current per channel, increases the bulk values of particle densities and shifts significantly the position of phase boundaries, although the overall topology of the phase diagram is preserved.

There are several extensions of the original two-channel TASEP that will be interesting to investigate.  In this paper we considered only the case with symmetric coupling, i.e., for any particle the probability of jumping to a different lattice chain is independent of the channel. It will be interesting to analyze the asymmetric coupling, where particles in the first channel could hop to the second channel with the rate $w_{1}$, while the opposite motion has the rate of $w_{2}$, and $w_{1} \neq w_{2}$. Another more interesting, although  more complex, case  is the problem of non-uniform couplings between the channels. It seems reasonable to suggest that our method of combining simple approximate theory with computer Monte Carlo simulations is a promising approach to study these complex non-equilibrium one-dimensional problems.

\ack

The support from the Camille and Henry Dreyfus New Faculty Awards Program (under Grant No. NF-00-056), from the Welch Foundation (under Grant No. C-1559), and from the US National Science Foundation through the grant CHE-0237105 is gratefully acknowledged. ABK also thanks T. Chou for valuable discussions.

\end{document}